\documentclass[a4paper,11pt]{article}
\usepackage{pos}

\title{The effect of light sea quark symmetry breaking on
	polarized nucleus and sum rules}

\author*[a]{Fatemeh Arbabifar}
\author[b]{Shahin Atashbar Tehrani}

\affiliation[a]{Department of Physics, Nasibeh campus, Farhangian University, \\ Tehran,Iran}

\affiliation[b]{School of Particles and Accelerators, Institute for Research in Fundamental Sciences (IPM),\\
P.O.Box 19395-5531, Tehran, Iran}

\emailAdd{F.Arbabifar@cfu.ac.ir}
\emailAdd{Atashbar@ipm.ir}

\abstract{The polarized structure functions of $^3He$ and $^3H$ nuclei are calculated in NLO approximation, considering
	and disregarding the light sea quark symmetry breaking. We employ the polarized structure function of the
	nucleons within the nucleus extracted from our two recent analysis on polarized DIS data and on polarized
	DIS+SIDIS data. Since the data of the second analysis cover a bigger range of Bjorken variable, both SU(2)
	and SU(3) symmetry breaking is considered within the analysis. Then we calculate and compare the polarized
	structures of nuclei extracted from both scenarios. Also the Bjorken and ELT sum rules are calculated using
	the moments of structure functions. The results are compared with experimental data and the differences are investigated.
}

\FullConference{%
  41st International Conference on High Energy physics - ICHEP2022\\
  6-13 July, 2022\\
  Bologna, Italy
}

\begin{document}
\maketitle
\section{Introduction}
Determination of polarized structure functions of nucleons and nuclei from  experimental data plays an important role at high energy physics. For instance it provides precise information about the distribution of the total spin of nucleons and nuclei on their content partons. In theoretical framework the next to leading order approximation of polarized structure function $x g_1(x, Q^2)$ in Bjorken $x$ space can be constructed based on the method of Jacobi polynomial $ \Theta_n^{\alpha, \beta}(x) $ expansion \cite{Nematollahi:2021ynm,Khanpour:2017fey,Mirjalili:2022cal}
\begin{eqnarray}\label{eq6}
	x g_1(x, Q^2) & = & x^{\beta}(1 - x)^{\alpha} \, \sum_{n = 0}^{\rm N_{max}} \, \Theta_n^{\alpha, \beta}(x)   
	\times  \sum_{j = 0}^n \, c_j^{(n)}{(\alpha, \beta)} \, {\cal M}[x g_1, j + 2](Q^2) \,,
\end{eqnarray}
where $ {\cal M}[x g_1, j + 2]$ is the moment of polarized structure function, $c_j^{(n)}$ is expansion coefficient, $N_{max}=9$ , and $\alpha=3$, $\beta=0.5$.\\
The nuclei of Helium-3 $^3He$ and Tritium $^3H$ are trivalent light nuclei that consist of two protons plus one neutron and two neutrons plus one proton respectively. The higher states $S^\prime$ and $D$ of their wave function can be found in a more realistic definition than the ground state level. Therefore, the calculation of their polarized structure function is performed utilizing the contribution of polarized structure function of proton and neutron plus the spin-dependent nucleon light-cone momentum
distributions $\Delta f^N_{^3{\mathrm He}}$ and $\Delta f^N_{^3{\mathrm He}}$ ~\cite{Yazdanpanah:2009zz}:
\begin{eqnarray}\label{g1Hesh}
	&& g_1^{^3{\mathrm He}}=\int_x^3 \frac{dy}{y} \Delta f^n_{^3{\mathrm He}}(y) g_1^n(x/y)
	+2 \int_x^3 \frac{dy}{y} \Delta f^p_{^3{\mathrm He}}(y) g_1^p(x/y)- 0.014 \Big( g_1^p(x) - 4 g_1^n(x)\Big)  ,\\
	&&g_1^{^3{\mathrm H}} = 2 \int_x^3 \frac{dy}{y} \Delta f^n_{^3{\mathrm H}}(y) g_1^p(x/y)
	+ \int_x^3 \frac{dy}{y} \Delta f^p_{^3{\mathrm H}}(y) g_1^n (x/y)+ 0.014 \Big( g_1^p(x) - 4 g_1^n (x) \Big).  
\end{eqnarray}

In the analysis of NAAMY21~\cite{Nematollahi:2021ynm}, the inclusive polarized DIS\footnote{Deep Inelastic Scattering} data are used and the polarized parton distribution functions of nucleons are extracted in NLO approximation. In the analysis of AKS14~\cite{Arbabifar:2013tma}, the experimental data of inclusive and semi-inclusive polarized DIS are applied and both flavor SU(2) and SU(3) symmetry breaking are considered. In the current study, the results of both of theses phenomenological models are used and the polarized structure functions of  Helium-3 ($^3He$) and Tritium ($^3H$) are extracted and compared.
\section{Sum rules and results}
Bjorken sum rule is effective to calculate the difference of the ﬁrst moment of proton and neutron
polarized structure functions considering nucleus constraint and disregarding it. So the ratio $\eta$ and its value from refs. \cite{Budick:1991zb,ParticleDataGroup:2016lqr} can be introduced as
\begin{eqnarray}\label{eq:sumratio}
	\eta &\equiv& \frac{g_A|_{triton}}{g_A} =  \frac{\int_0^3[g_1^{^3{\mathrm H}}(x,Q^2)-g_1^{^3{\mathrm He}}(x,Q^2)]dx} {\int_0^1[g_1^p(x,Q^2)-g_1^n(x,Q^2)]dx} = 0.956 \pm 0.004 \,. \nonumber \\
\end{eqnarray}
We obtained a value of 0.936376 from NAAMY21~\cite{Nematollahi:2021ynm} results and 0.954312 from AKS14~\cite{Arbabifar:2013tma} results. These values show that $\eta$ extracted from the analysis with symmetry breaking consideration is closer to the value of Bjorken sum rule.\\
The Efremov-Leader-Teryaev (ELT) sum rule is obtained via integrating the valence part of longitudinal and transverse structure functions over $x$ in the limit of $m_q\rightarrow 0$ \cite{blum:1997}
\begin{eqnarray}\label{eq:ELT}
	\int_0^1x[g_1^{V}(x)+2g_2^{V}(x)]dx=0~,
\end{eqnarray}
where $g_{1(2)}^{V}$ presents the valence quark contributions to $g_{1(2)}$. 
Considering the symmetry of light sea quarks, ELT sum rule can be written as
\begin{eqnarray}\label{eq:ELTsym}
	\int_0^1x[g_1^{p}(x)-g_1^n+2(g_2^{p}(x)-g_2^n(x)]dx=0~.
\end{eqnarray}
\begin{figure}[!htb]
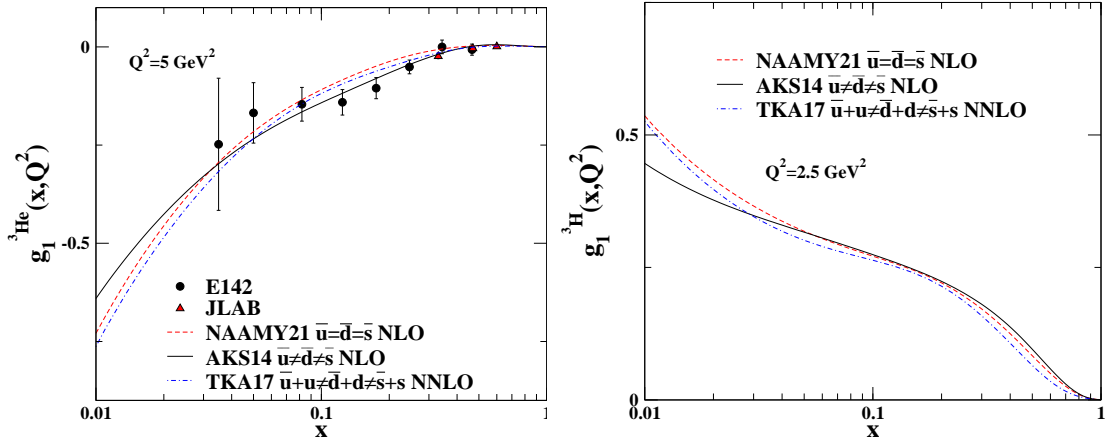

	\centering
	\includegraphics[width=0.32\textheight]{g1He3-2.eps}
	\includegraphics[width=0.32\textheight]{g1H3-3.eps}
	\caption{{\small $g_1^{^3He}$ polarized structure functions in NLO and NNLO approximation from Refs. \cite{Nematollahi:2021ynm,Arbabifar:2013tma} and \cite{Khanpour:2017fey} compared with E142~\cite{E142:1996thl} and JLAB~\cite{JeffersonLabHallA:2004tea} experimental data (left), $g_1^{^3H}$ polarized structure functions in NLO and NNLO approximation from Refs. \cite{Nematollahi:2021ynm,Arbabifar:2013tma} and \cite{Khanpour:2017fey}  (right).  \label{fig:fig1}}}
\end{figure}

\begin{figure}[!htb]
	\begin{center}
		\includegraphics[clip,width=0.42\textwidth]{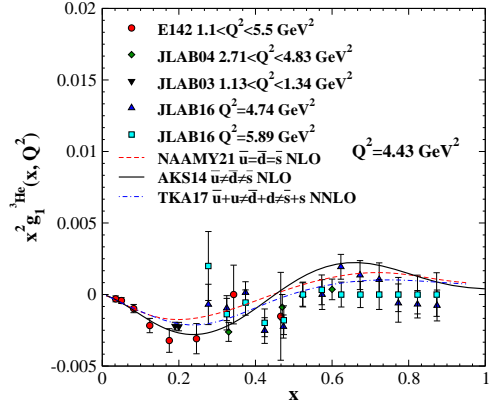}
		\caption{{\small $x^2g_1^{^3He}$ polarized structure functions in NLO and NNLO approximation from Refs. \cite{Nematollahi:2021ynm,Arbabifar:2013tma} and \cite{Khanpour:2017fey} compared with E142~\cite{E142:1996thl}, JLAB04~\cite{JeffersonLabHallA:2004tea}, JLAB03~\cite{jlab:WilliamandMary} and JLAB16~\cite{JeffersonLabHallA:2016neg} experimental data. \label{fig:fig2}}}
	\end{center}
\end{figure}
When SU(2) and SU(3) symmetry breaking are considered, the ELT sum rule is directly derived from Eq.~\ref{eq:ELT}. The value of above equation is obtained $-0.011\pm 0.008$ from E155\cite{E155:2003} analysis at $Q^2=5~GeV^2$. This value is obtained $0.01017\pm0.00004$ from NAAMY21 and $-0.030763\pm0.0004071$ from AKS14. It seems that symmetry breaking makes the extracted value negative and closer to E155 obtained value.\\
Figure~\ref{fig:fig1} shows the polarized structure functions of $^3He$ and $^3H$ at NLO and NNLO obtained from NAAMY21~\cite{Nematollahi:2021ynm} and TKA17~\cite{Khanpour:2017fey} which concern the symmetry of light sea quarks in comparison with the results of AKS14 \cite{Arbabifar:2013tma} model at NLO  and also with E142~\cite{E142:1996thl}and JLAB~\cite{JeffersonLabHallA:2004tea} experimental data. It can be seen that the curve obtained from AKS14 is in better agreement with the experimental data in the most areas of $x$ variable. In fact, AKS14, NAAMY21 and TKA17 curves respectively pass 11, 6 and 9 data points or their error bar for $g_1^{^3He}$. There is no experimental data for $g_1^{^3H}$ to compare with curves but the difference between the curves is noticeable. Figure~ \ref{fig:fig2} shows the comparison of $x^2g_1^{^3He}$ polarized structure function with E142~\cite{E142:1996thl}, JLAB04~\cite{JeffersonLabHallA:2004tea}, JLAB03~\cite{jlab:WilliamandMary} and JLAB16~\cite{JeffersonLabHallA:2016neg} experimental data at different range of energies. It confirms that the curve from AKS14 model behaves better than other curves in the region of $0.1\leq x \leq 0.4$. It comes from the fact that the peaks and valleys of polarized parton distribution of light sea quarks of AKS14 model are mostly in $x \leq 0.4$ , so their effects in these areas is more noticeable.
 At higher region of $x$, NAAMY21 and TKA17 curves look closer to the experimental data and their error bars.

\end{document}